\documentclass[a4paper,11pt]{iopart}
\usepackage{graphicx,sidecap}
\usepackage{bm}
\usepackage{amssymb}
\usepackage{cite}
\usepackage{amsfonts}
\usepackage{mathrsfs}
\usepackage{xcolor}
\usepackage{hyperref}
\usepackage{enumerate}
\usepackage{verbatim}
\usepackage[margin=30pt, bf, font=footnotesize, center, justification=justified]{caption}[2004/07/16]
\usepackage{multirow}
\usepackage{soul}

\usepackage{tikz}
\usetikzlibrary{decorations.pathmorphing}
\usetikzlibrary{backgrounds}

\hypersetup{colorlinks,bookmarksopen,bookmarksnumbered,
citecolor=red,
linkcolor=blue,
pdfstartview=green,
urlcolor=teal}

\begin{document}

\title[Anomalous transport in FPUT]{Anomalous transport in the 
Fermi-Pasta-Ulam-Tsingou model: a review and open problems}

\author{Stefano Lepri$^{1,2}$, Roberto Livi$^{3,1,2}$,
Antonio Politi$^{1,4}$}

\address{$^{1}$ Istituto dei Sistemi Complessi, Consiglio Nazionale
delle Ricerche, via Madonna del Piano 10, 50019 Sesto Fiorentino, Italy}
\address{$^{2}$ Istituto Nazionale di Fisica Nucleare, Sezione di
    Firenze, via G. Sansone 1, 50019 Sesto Fiorentino, Italy}
\address{$^{3}$ Dipartimento di Fisica e Astronomia, Universit\`a di Firenze, Via G. Sansone 1, I-50019
Sesto Fiorentino, Italy}
\address{$^{4}$ Institute for Complex Systems and Mathematical Biology, University of Aberdeen, Aberdeen AB24 3UE, United Kingdom}

\eads{\mailto{\mailto{stefano.lepri@cnr.it}, roberto.livi@unifi.it},
  \mailto{a.politi@abdn.ac.uk}}


\begin{abstract}
This review provides an up-to-date account of energy transport in Fermi-Pasta-Ulam-Tsingou (FPUT) chains, a key testbed for nonequilibrium statistical physics. We discuss the transition from the historical puzzle of thermalization to the discovery of anomalous heat transport, where the effective thermal conductivity $\kappa$ diverges with system size $L$ as $\kappa \propto L^\delta$. 
The article clarifies the distinction between two universality classes: the FPUT-$\alpha \beta$ model, characterized by $\delta = 1/3$ and linked to Kardar-Parisi-Zhang (KPZ) physics, and the symmetric FPUT-$\beta$ model, where numerical and theoretical evidence support $\delta = 2/5$. We investigate how finite-size effects - unavoidably induced by the thermostatting protocols -
can disguise the asymptotic scaling. Additionally, we analyze the role of conservative noise in preserving hydrodynamic properties and examine how proximity to integrable limits leads to long-lived quasi-particles and, thereby, to
diffusive regimes over intermediate spatial scales.
\end{abstract}
\vspace{10pt}
\noindent \textit{Keywords}: FPUT, anomalous transport, heat conduction, KPZ universality, nonlinear chains.

\section{Introduction}
 \label{intro}
As testified by the contributions to this Special Issue, the FPUT numerical experiments have been
and still are a fruitful source of inspiration for many mathematical and physical questions.  Throughout the years, The FPUT model proved to be a simple
yet rich system to investigate various nonlinear phenomena as well
as the basics of  relaxation and transport mechanisms in 
interacting many-particle classical systems \cite{campbell2005introduction ,gallavotti2007fermi}. 

One of the surprising ``little discoveries" encountered therein is the \textit{anomaly}
of energy transport, meaning that 
usual Fourier's law of heat conduction breaks down \cite{LLP97}.     
For almost three decades from these finding, a considerable amount of 
research was devoted to this topic.  
The basics concepts and results have been summarized already 
in the first review articles on the topic \cite{LLP03,DHARREV,lepri2016heat}, 
and in the monograph \cite{Lepri2016}.  
A more updated account, including 
applications to condensed matter and nanosystems, 
can be found in \cite{benenti2020anomalous,RNC}.

Many aspects have been clarified and the problem has been put in 
a broader context of nonequilibrium statistical physics. 
In particular, universal (i.e. model independent) scaling laws 
have been strongly advocated and traced back to the connection 
between the microscopic FPUT dynamics an the Kardar-Parisi-Zhang 
physics \cite{Spohn2014}.
This connection demonstrates vividly how  universal features
found for simple models like FPUT provided important insights
on technologically relevant issues,  like nanoscale heat transport
and conversion \cite{RNC}.

However, to our own surprise, there are still open issues.  
This review article attempts to provide an up-to-date account of energy transport 
focused on the FPUT-chains. To illustrate the problem status
we also provide new simulation data that may hopefully stimulate 
further research.  
 
As an appetizer, in Section \ref{sec:anomal}, we start recalling the historical pathway that made the FPUT model a
key testbed for the study of heat transport in classical nonlinear chains, which eventually led to the 
unexpected evidence of a divergent heat conductivity $\kappa$ with the system size $L$.
Basic definitions of the FPUT model and of the relevant observables are then presented together with
a brief summary of the overall scenario.
In fact, nowadays, there is a wide consensus that for the so-called FPUT-$\alpha \beta$ model 
(with cubic and quartic nonlinerities), $\kappa \simeq L^{\delta}$ with $\delta = \frac{1}{3}$.
The FPUT-$\beta$ model (with quartic nonlinearity only), instead, belongs to a different universality class:
the value of the exponent has been at length debated, but eventually numerical experiments and 
theoretical arguments provide compelling evidence for $\delta =  \frac{2}{5}$.

 In Section~\ref{finsize} we discuss how finite-size effects may disguise the evidence of the asymptotic scaling
 behavior. First, we investigate the role of the length of the thermostatted portion of the chain
(via a Langevin dynamics), eventually finding that single-particle thermostats induce less appreciable
finite-size effects. Then, we focus on the role of coupling strength (in this case using thermostats that
randomize the particle velocity according to a suitable equilibrium distribution), confirming that
{\it intermediate} strengths provide optimal estimates.

Section \ref{sec:equil} deals with equilibrium simulations. We first discuss the relationship with
the KPZ evolution of rough interfaces, offering a direct evidence via to the reconstruction
of the associated sound modes. Then, we illustrate the differences between  
the above mentioned universality classes through the study of the relaxation of
the the correlation functions of sound and heat modes in the FPUT chains. 
The numerical results turn out to be in agreement with the exponent $\delta=\frac{2}{5}$
observed in the FPUT-$\beta$ model, and inevitably inconsistent with fluctuating hydrodynamics.

In Section~\ref{consnoise} we discuss the role of conservative noise. It is known that
the inclusion of conservative noise in chains of harmonic oscillators induces an anomalous
heat conductivity, characterized by an exponent $\delta=\frac{1}{2}$~\cite{BBO06,basile2016thermal}. 
In order to test the robustness of the universality class, we add conservative noise to 
an FPUT-$\alpha \beta$ chain. As a result, we show that, contrary to a previous conjecture, 
the presence of conservative noise does not modify the hydrodynamic properties, although 
the convergence to the exponent $\delta = \frac{1}{3}$ occurs for larger system sizes.

In Section \ref{sec:transint} we discuss how the closeness of the integrable
limit affects thermal transport, and review some unexpected features
that are caused by the presence of long-lived quasi-particles.
This implies a special type of finite-size effects occurring in FPUT-like models, 
when they are close to an integrable limit. For instance, the Toda-chain can be viewed as
 the integrable model closer to the FPUT-$\alpha \beta$ model, while the low-energy limit of the FPUT-$\beta$ model
 corresponds to the integrable harmonic chain. This simple remark suggests that these two models should exhibit
 quite different hydrodynamic behaviors also when considered as perturbations of integrable models. 
 
 Final remarks and future perspectives are contained in the Conclusions.

\section{Generalities about energy transport in the FPUT-model} 
\label{sec:anomal}

\subsection{A historical perspective}
In the thirties of the XX century, P. Debye was the first scientist, to our knowledge, to speculate about the possibility 
of studying the problem of heat (energy) transport in a microscopic model of matter. He realized that this
problem cannot be properly modeled by a harmonic crystal, because it amounts to 
a gas of non-interacting quasi-particles, the harmonic waves (that we now call phonons), 
which propagate ballistically through the lattice at sound speed. Since no scattering mechanism
is present in this model, perturbations propagate ballistically, making the solid a heat superconductor.
A rigorous solution of energy transport in a harmonic chain in contact with two thermal reservoirs 
at different temperatures $T_+$ and $T_-$, was given a few decades later~\cite{rieder1967}. 
The total heat flux was found to be proportional to the temperature difference, independently of the
chain length $L$, instead of being inversely proportional as expected from the Fourier's law.
Moreover, it was found that the temperature profile is practically flat at the average value $T= (T_+ +T_-)/2$, 
accompanied by two discontinuities at the boundaries.
 
P. Debye suggested that in order to recover a diffusive transport one should add two further ingredients 
in the lattice model, namely nonlinear interactions and disorder.  Almost two decades later this seminal 
conjecture presumably inspired  E. Fermi and his collaborators, J. Pasta, S. Ulam and M. Tsingou, 
who investigated energy thermalization in a lattice of nonlinearly coupled oscillators, a system nowadays
known as the Femi-Pasta-Ulam-Tsingou (FPUT)-model
(for a recent historical account see \cite{lepri2023fermi}).  They 
could access the biggest computer facility of the time, the MANIAC digital computer in Los Alamos, designed 
by J. von Neumann to perform calculations for the Manhattan Project.  In fact, the basic intuition of Fermi was 
that the presence of nonlinear interactions could allow any out-of-equilibrium initial condition (e.g. a packet of low 
frequency harmonic modes)  to relax spontaneously to  thermodynamic equilibrium, signaled by energy 
equipartition among all the Fourier modes. Contrary to his expectations, 
the numerical simulations showed that the energy of
the out-of-equilibrium initial condition, initially transferred to some close unexcited modes, 
was later mostly returning to the initial state, following a quasi-periodic recurrent dynamics. 
Nowadays, it is well known 
that in the low-energy limit of the model explored by Fermi and coworkers in their simulations,
the almost recurrent dynamics is due to the presence of soliton-like waves \cite{kruszab1967}, which survive over extremely
long time scales  .
 
Fermi was seriously puzzled by this outcome, but unfortunately he had no time for further investigations. 
In fact, in May 1955, six months after
he passed away, an account of  these numerical simulations appeared as an internal Los Alamos report \cite{Fermi1955}.
It is worth pointing out that the quasi-periodic recurrence to the initial states did not only challenge the 
spontaneous evolution to thermodynamic equilibrium expected by Fermi, but it was also incompatible with the presence of any diffusive mechanism 
for energy transport in the FPUT-model.  
Making a long story short, thirteen years later, energy equipartition in the FPUT-model was eventually observed in numerical
experiments performed at larger energy values and
using significantly more powerful computers~\cite{Chirikov}.  

For what concerns energy transport in the FPUT-model, some attempts of studying thermal conductivity 
were essentially not conclusive, due to the limited computational power \cite{Jackson, Nakazawa}. 
Later, the numerical evidence of Fourier's law in a simplified chaotic model~\cite{Casati} 
led to the conjecture that a well established chaotic dynamics is a necessary and sufficient condition
to induce a diffusive energy spread.

At the end of last century, when computer facilities allowed for larger-scale simulations,  the problem of heat transport 
in the FPUT-model was reconsidered  \cite{LLP97,Lepri98a}. These studies provided the first evidence of anomalous energy transport (more 
precisely, a power-law  divergence of the  heat conductivity in the thermodynamic limit) in low-dimensional  systems. These unexpected results
attracted a renewed interest on the problem, while unveiling also the possibility of experimental verification of anomalous thermal conductivity  
in nano-materials, e.g. nanotubes, polymers, atomic chains  and  graphene layers. A wide literature has been devoted to more and more refined
studies: rather than reporting too large a list of references we address the reader to some review papers
 \cite{LLP03,DHARREV,Lepri2016,RNC,Brunico}.

\subsection{Notations and observables}

The FPUT-model is a chain of $N$ classical point--like particles with equal mass $m$, whose positions, $q_n$, and
momenta, $p_n = {\dot q}_n$, are  the canonical coordinates of the Hamiltonian
\begin{equation}
{H} = \sum_{n=1}^N \left[{p_n^2\over 2m} +
V(q_{n+1}-q_{n})\right] \quad .
\label{hami}
\end{equation}
The potential $V(r)$ accounts for the nearest-neighbor interactions between consecutive particles
and consists of the fourth-order Taylor series expansion of any phenomenological 
nonlinear molecular potential,
\begin{equation}
V(r)\;=\; \frac{k_2}{2}\, r^2+ \frac{k_3}{3}\, r^3 + 
\frac{k_4}{4}\, r^4 \quad .
\label{fpu}
\end{equation} 
The corresponding evolution equations are
\begin{equation} 
m{\ddot q}_n = - F(r_{n}) + F(r_{n-1})  \quad , \quad n=1,\ldots,N \, ,
\label{eqmot} 
\end{equation}  
where $r_n=q_{n+1}-q_n$, $F(r)=- V'(r)$, and the prime denotes a derivative with respect to the argument.
The parameter $k_2$ in front of the quadratic term is the harmonic coupling constant. In the 
original work by Fermi and coworkers \cite{Fermi1955}, the couplings 
$k_3$ and $k_4$ were denoted by $\alpha$ and $\beta$, respectively. This is why this model
was named the ``FPUT-$\alpha \beta$'' model, while its version with $k_3=0$ 
is known as  the ``FPUT-$\beta$'' model. 


In the constant-volume thermodynamic ensemble, there are three conserved quantities, namely
the total length $L$, total momentum $P$, and total energy $E$, which can all be expressed in
terms of local variables,
\begin{eqnarray}
&& L =  \sum_{n=1}^N r_n \nonumber \\
&& P=   \sum_{N=1} m \dot q_n \equiv \sum_{n=1}^N p_n  \label{eq:localv}\\
&& E = \sum_{n=1}^N \left \{
    {p_n^2\over 2m} + \frac12 \bigg[ V(r_n) +
    V(r_{n-1}) \bigg] \right \} \equiv \sum_{n=1}^N e_n \nonumber \quad ,
\end{eqnarray}
where $r_n$, $p_n$, and $e_n$ are the microscopic elements at basis of fluctuating thermodynamics.

Finally, in the context of nonequilibrium stationary states and within the Lagrangian framework,
the energy flux can be defined as
\begin{equation}
j_n = \frac{1}{2} ({\dot q}_{n+1} + {\dot q}_n)F(r_n) \,.
      \label{hf2}
\end{equation}
As discussed in Ref.~\cite{MejiaMonasterio19}, an Eulerian exact definition of the flux
is more involved, as it requires separately dealing with the conductive and the convective component.
Here, we shall always make use of the simple Lagrangian definition, Eq.(\ref{hf2}).


\subsection{Anomalous transport in the FPUT-model}
\label{sec:signatures}

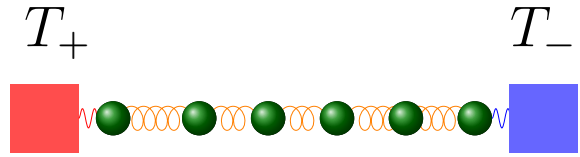
\begin{figure}[t]
\begin{center}

\resizebox{0.5\textwidth}{!}{%
 \begin{tikzpicture}
\fill[red!70!white] (-4,-3) rectangle (-2,-1) ;
\draw[thick,red,decorate,decoration={snake,aspect=0.5,amplitude=8}] (-2.,-2) -- (-1,-2);
\draw[thick,orange,decorate,decoration={coil,aspect=0.7,amplitude=10}] (-.8,-2) -- (1.5,-2);
\draw[thick,orange,decorate,decoration={coil,aspect=0.7,amplitude=10}] (1.8,-2) -- (3.5,-2);
\draw[thick,orange,decorate,decoration={coil,aspect=0.7,amplitude=10}] (3.8,-2) -- (5.5,-2);
\draw[thick,orange,decorate,decoration={coil,aspect=0.7,amplitude=10}] (5.8,-2) -- (9.5,-2);
\fill[ball color= green!50!black] (-1.,-2) circle (.5);
\fill[ball color= green!50!black] (1.5,-2) circle (.5);
\fill[ball color= green!50!black] (3.5,-2) circle (.5);
\fill[ball color= green!50!black] (5.5,-2) circle (.5);
\fill[ball color= green!50!black] (7.5,-2) circle (.5);
\fill[ball color= green!50!black] (9.5,-2) circle (.5);\draw[thick,blue,decorate,decoration={snake,aspect=0.5,amplitude=8}] (10,-2) -- (11.5,-2);
\fill[blue!60!white] (10.5,-3) rectangle (12.5,-1);
\draw (-2.6,0.5) node[scale=4]{$T_+$};
\draw (11.5,0.5) node[scale=4]{$T_-$};
\end{tikzpicture}
}%
\end{center}
\caption{A one-dimensional chain of coupled oscillators
interacting with two thermal reservoirs ad different temperatures $T_+$ and $T_-$.  Actual implementation of the reservoirs can be either stochastic 
(Langevin) or deterministic (via e.g. isokinetic thermostat), see \cite{LLP03}
for details.}
\label{fig:baths} 
\end{figure}

Heat transport is often investigated by referring to the conceptual setup sketched in
Fig.~\ref{fig:baths}, where the chain of oscillators is in stationary out-of-equilibrium conditions, 
ensured by two thermal reservoirs acting at its boundaries.  

The effective heat conductivity $\kappa (L)$ can be determined by exploiting the formula
\begin{equation}
\kappa (L) = - \frac{J (L)}{\nabla T}  \quad\quad ,
    \label{Fourier}
\end{equation}
where $J(L)$ is the stationary energy current flowing through the chain.
In the large system-size limit (i.e for $L\to \infty$ ), anomalous transport is signaled
by a divergence of the effective conductivity \cite{LLP97} ,
\begin{equation}
\kappa(L) \;\propto\; L^\delta \quad\quad ,
    \label{diverge}
\end{equation}
where $0 < \delta < 1$.  Numerical 
studies indicate that while the actual value $J(L)$ of the flux does depend on the selection of boundary
conditions (e.g. free vs. fixed), the divergence rate is universal.

\begin{figure}[ht!]
\centering \includegraphics [width=0.7\textwidth,clip]{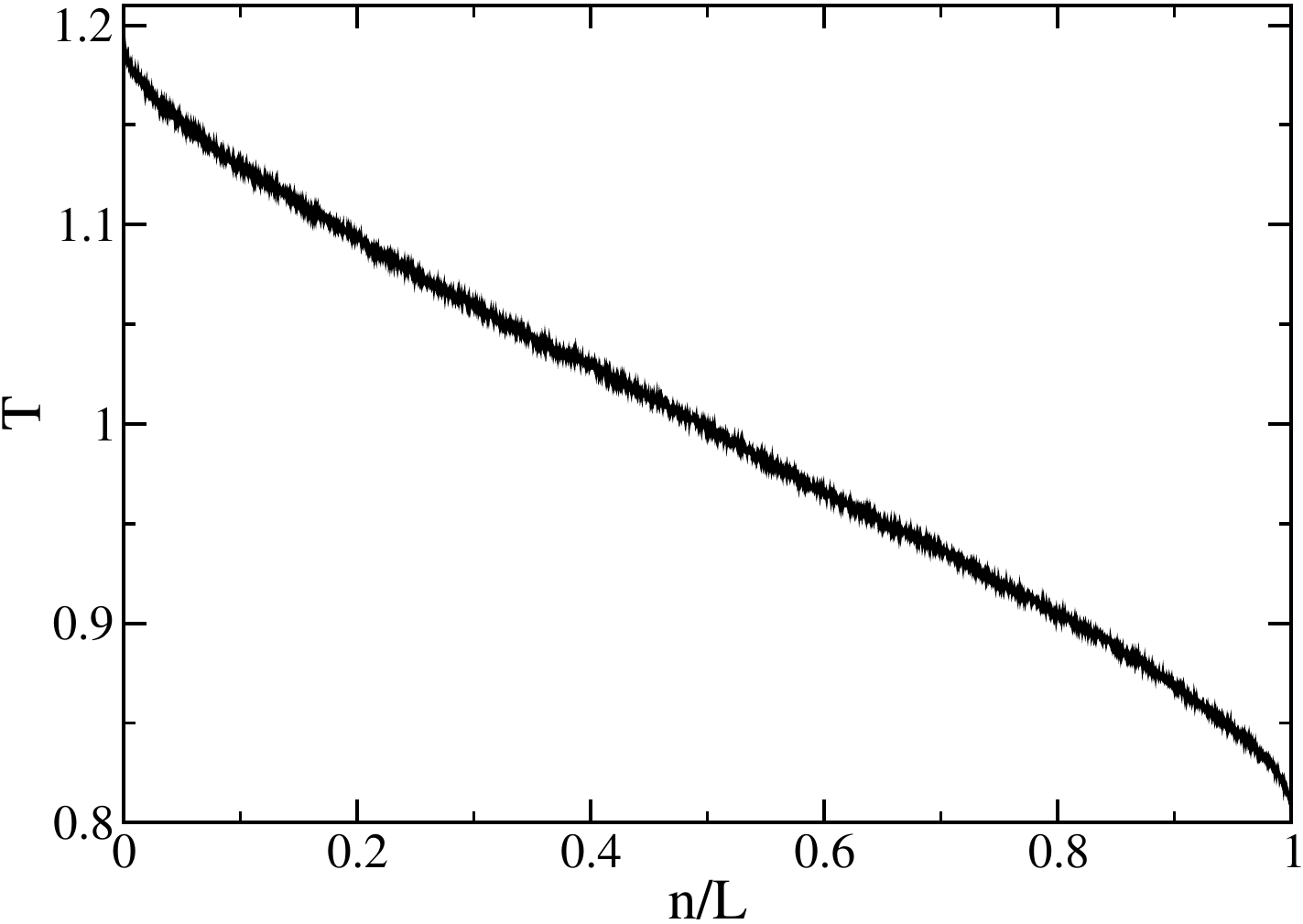}
\caption{Kinetic temperature profile 
$\langle p_n^2\rangle$ for an FPUT-$\beta$ chain with 
$L=8192$ particles and thermostats sets at
$T_+ = 1.2$, $T_-=0.8$, respectively.
}
\label{fig:prof_temp}
\end{figure}

Additional indirect evidence of anomalous transport is offered by the nonlinear 
${\it S-}$ shaped kinetic temperature profile 
(see Fig.~\ref{fig:prof_temp}). This arises even when the temperature difference between the two thermostats is very small, and cannot thus be 
simply explained as an effect of temperature-dependent conductivity.
Instead, this is a clear manifestation of long-range correlations. 
In fact, the temperature field is a solution of 
a fractional heat equation with suitable boundary conditions \cite{Lepri2011,Kundu2019,Dhar2019}.

Alternatively, equilibrium simulations can be performed, thereby focusing  on the fluctuations of the total energy 
current $J = \sum_{n=1}^N \, j_n$\footnote{This can be done by either
setting $T_+ = T_-$ ({\sl canonical} setup), or performing microcanonical simulations, preferably
using periodic boundary conditions to minimize finite-size effects.}.
In fact, from linear response theory~\cite{kubo1985}, the temporal correlation function of the flux 
is expected to decay as
\begin{equation}
\langle J(t)J(0)\rangle \propto t^{-(1-\delta)}
\label{longtail}
\end{equation}
where $\delta$ is again the exponent ruling the divergence of heat conductivity (see \cite{LLP03}).
The correctness of this prediction has been tested and confirmed in many numerical experiments~\cite{LLP03,DHARREV,Lepri2016}.

Finally, equivalent evidence of anomalous transport can also be found by studying the diffusion of 
localized energy perturbations of equilibrium states. The variance of the perturbation indeed 
grows as~\cite{Denisov03,Cipriani05}
\begin{equation}
\sigma^2(t)  \propto t^{1+\delta}
\end{equation}
i.e., the evolution is superdiffusive when $\delta>0$.

In the last decades, several independent theoretical studies have been performed to
determine the behavior of generic models where total length, total momentum and total energy are conserved.
They are based on: (i) renormalization-group arguments \cite{NR02}; (ii) mode-coupling theory \cite{Lepri98c};
(iii) fluctuating hydrodynamics~\cite{VanBeijeren2012,Spohn2014}.
As a result, there is a general consensus that $\delta=\frac{1}{3}$: a prediction substantially
confirmed by numerical simulations including those of the FPUT-$\alpha \beta$ model.

\begin{figure}[ht!]
\centering \includegraphics [width=0.45\textwidth,clip]{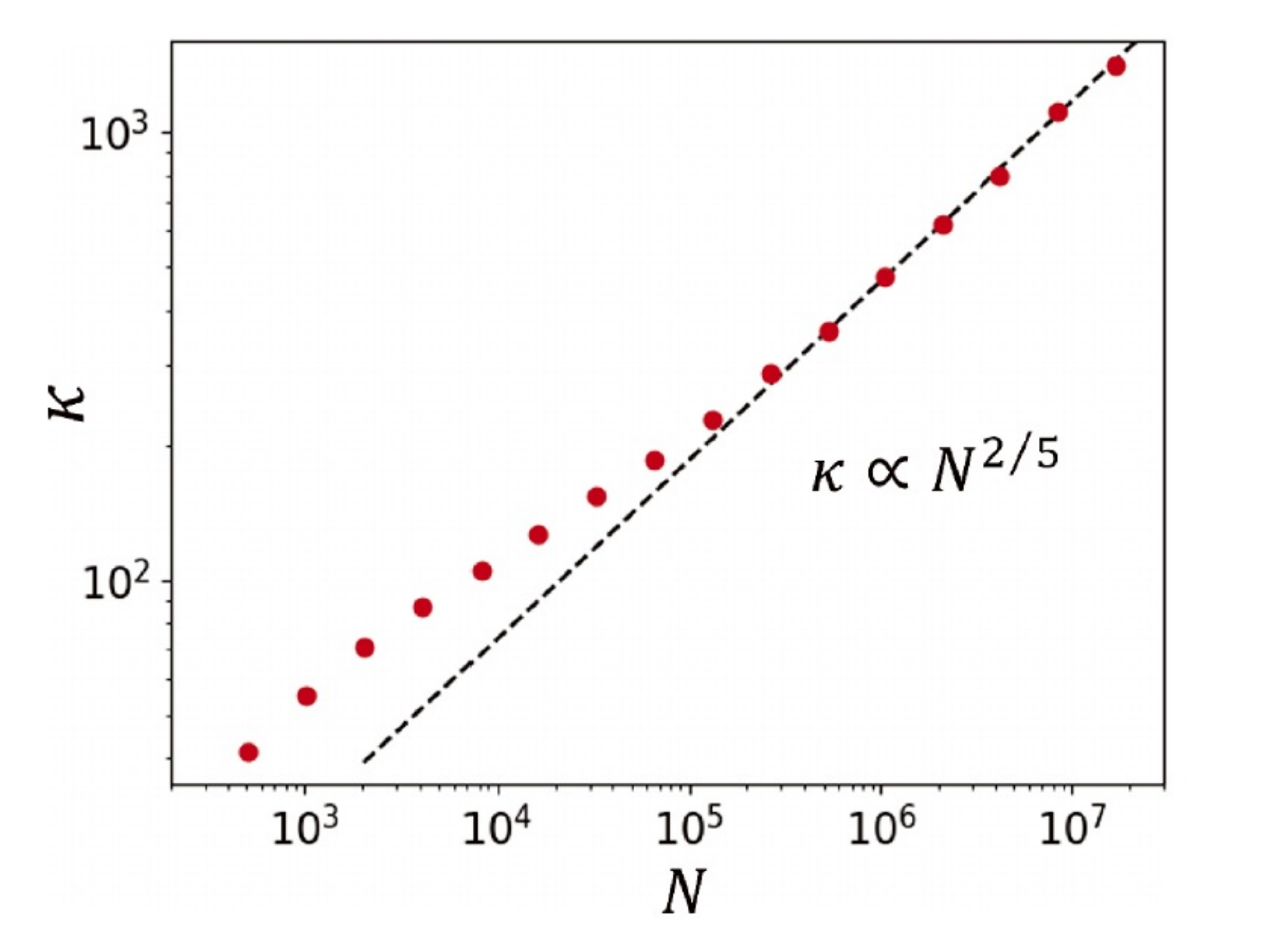}
\includegraphics [width=0.45\textwidth,clip]{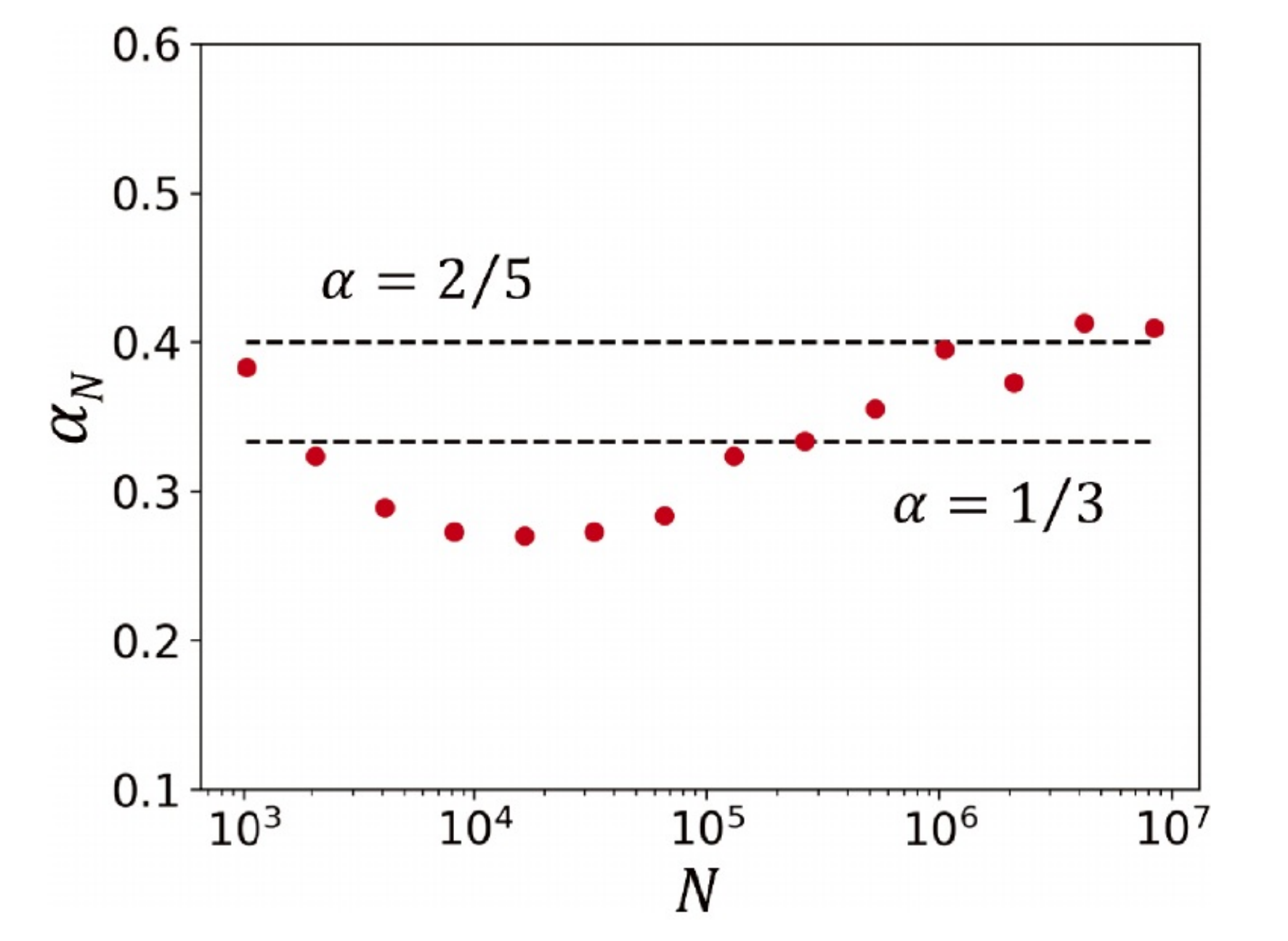}
\caption{Left: Thermal conductivity of FPUT-$\beta$
versus  lattice size $N$.  (denoted by $L$here);   Langevin heat bath at 
temperatures  1.2 and 0.8.  Dashed line represents the power
law $N^{2/5}$.  Right: logarithmic derivative
versus lattice size $N$. \textcopyright (2024),  The Physical Society of Japan, reprinted with permission from Ref.\cite{takatsu2024large}.
}
\label{fig:longbeta}
\end{figure}

More debated is the special case of symmetric potentials, such as the FPUT-$\beta$ model.
While fluctuating hydrodynamics and mode-coupling theory predict $\delta=1/2$, kinetic approaches 
(see~\cite{Pereverzev2003,Nickel07,Lukkarinen2008} and Chapter~4 in~\cite{Lepri2016} for a detailed account)
and Boltzmann's kinetic equations~\cite{Dematteis} yield $\delta=2/5$.
Nonequilibrium simulations are consistent with this latter prediction 
\cite{Lepri03,Wang2011}.  Remarkably,  the recent extensive study
reporting data for chains up to length $L=2^{24}\approx 1.6 \times 10^7$ definitely support the $\frac{2}{5}$ prediction
as shown in Fig.~\ref{fig:longbeta}.
It is however worth mentioning that some systematic deviations occur 
in the case of purely quartic FPUT potential \cite{xiong2018observing}.

\section{Finite-size effects}
\label{finsize}
The problem of heat conduction can be numerically addressed by either simulating a stationary setup
and thereby estimating the resulting heat flux, or by performing equilibrium simulations and looking at the correlations
of the flux \cite{LLP03,DHARREV} which are related to transport
coefficient by Green-Kubo formula.  In both cases, a careful examination 
of the dependence of the results on the lattice size $L$ is mandatory
and actually one of the most challenging issues.

The equilibrium approach has the advantage that it does not require the introduction of external ingredients
(the heat baths) and is not affected by the associated boundary effects.
On the other hand, nonequilibrium methods usually allow for a clearer separation of finite-time and finite-size effects.
In fact, once the length of the system has been fixed, the simulation time is (conceptually) well identifiable
as the minimal time such that the statistical fluctuations of the average flux become sufficiently small.
A second pragmatic advantage is the computation of the flux, since it is nothing but the energy exchanged
with the thermal baths, while in the equilibrium setup it is necessary to rely on a proper bulk definition which
may, sometimes, deserve some care,
e.g. in the presence of next to nearest neighbor interactions.
One unavoidable drawback is the presence of  \textit{boundary or Kapitza resistance},  a well-known feature that induces temperature jumps 
close to the thermostatted regions \cite{Aoki01}.

In this section, we revisit the stationary state which emerges under the 
action of two external thermal baths (see Fig. \ref{fig:baths}). This is done with reference to the
FPUT-$\beta$ model, but the various features we are going to illustrate are very general and apply to
generic nonlinear models.

In the literature, two thermostatting schemes are typically considered.
The first consists in simulating the thermal bath via  Langevin equations. 
Here, the coupling strength is identified by the interaction frequency with the chain of particles.
The second approach consists in randomizing the velocity of the particles in contact with the thermostats. This is basically the Andersen thermostat method
well-known in molecular dynamics.
In this case, the time separation between consecutive resets plays the role of the coupling strength.
For completeness, one should also mention the Monte Carlo method, which in some cases proves rather effective
(typically in models where the temperature is not directly related 
to the velocity distribution such as for the DNLS equation \cite{Iubini2012,Iubini2013a}).

It is advisable to select an intermediate coupling strength for simulations 
of stationary nonequilibrium setups, the 
reason being that both in the limit of small and strong coupling, a relatively sizable temperature gap arises at the 
boundaries~\cite{LLP03}.
Perhaps more important, \textit{de facto}, the flux is smaller and therefore 
affected by relatively larger statistical fluctuations.
However, there are no comprehensive studies where the quality of different procedures is discussed especially
as a function of the system size.

In the following two sub-sections, we explore the role played by the number of thermostatted oscillators and by  the
coupling strength.

\subsection{Length of the thermostatted region}

To our knowledge, almost all numerical studies of a stationary non-equilibrium setup have been performed by assuming that one single site is coupled to the thermostat. 
This choice originates from the seminal 
work \cite{RLL67} for the harmonic
chain and it is mostly dictated by a reason of simplicity.
However, in the literature, some authors choose to couple several sites (say up to 20) under the assumption
of a smoother interaction and perhaps of a greater realism.
This is for instance the case for the most recent 
and extensive simulations of heat conductivity in the FPUT-$\beta$ model \cite{takatsu2024large}.

As a premise, it is worth noticing that in the thermodynamic limit the conductivity, or better the leading contribution
to the conductivity, is expected to be a bulk property, independent of the way the boundaries are treated.
In the presence of anomalous conductivity, this is not entirely true because, as implicitly recognized in  Ref. \cite{Delfini08c}
in the FPUT-$\beta$ chain, the conductivity, while scaling in the same way both for fixed and free b.c., turns out to be
quantitatively larger in the second case .
So, it is important to understand to what extent the estimated values depend also on the choice of the heat baths.

\begin{figure}[ht!]
\centering \includegraphics [width=0.6\textwidth,clip]{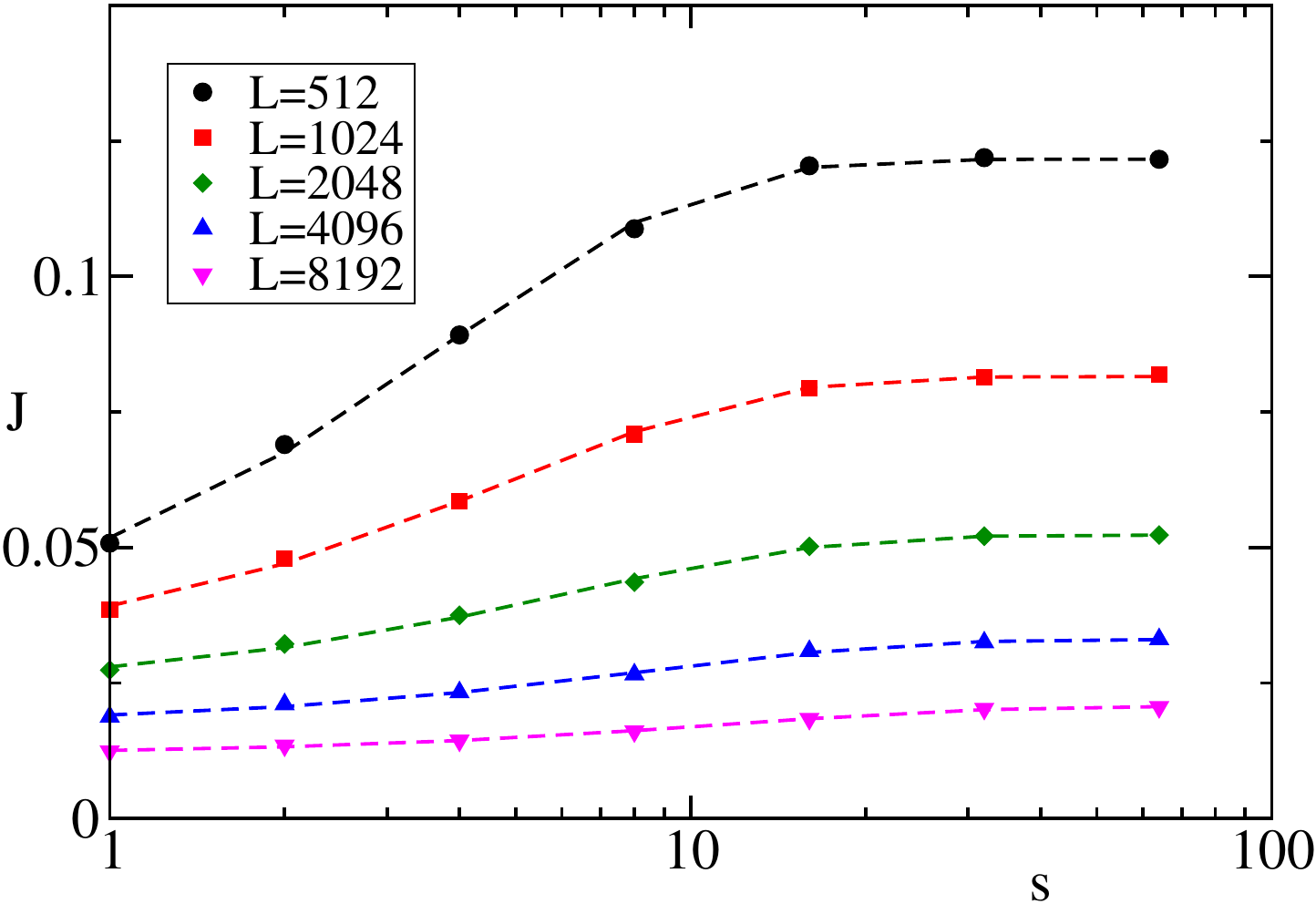}
\caption{Energy flux in the FPUT-$\beta$ model, versus the length $s$ of the thermostatted region, for different chain lengths.}
\label{fig:flussi_s}
\end{figure}

Here, we consider Langevin heat baths and, in order to be able to use symplectic algorithms, we implement them 
by first integrating the Hamiltonian equations over a time step ($=0.01$ units) and afterwards 
reducing the velocity by a given fraction and
simultaneously adding a stochastic term in such a way that fluctuation dissipation is satisfied for the given preassigned
temperature.
A chain of length $L+2s$ is considered such that the first and last $s$ particles interact as described above.
The coupling strength is fixed to $\gamma=0.3$, while the left and right temperatures are fixed equal to 2 and 0.5, 
respectively. We have determined the energy flux for $s=1$, 2, 4, 8, 16, 32, and 64, performing simulations for
$L=512$, 1024, 2048, 4096, and 8192.
The resulting values are presented in Fig.~\ref{fig:flussi_s}, where we see that, quite naturally, the flux increases
for increasing $s$ since the number of channels which exchange energy with the chain grows.
The saturation is a clear indication that the maximal transport capacity of the chain has been reached.

\begin{figure}[ht!]
\centering \includegraphics [width=0.6\textwidth,clip]{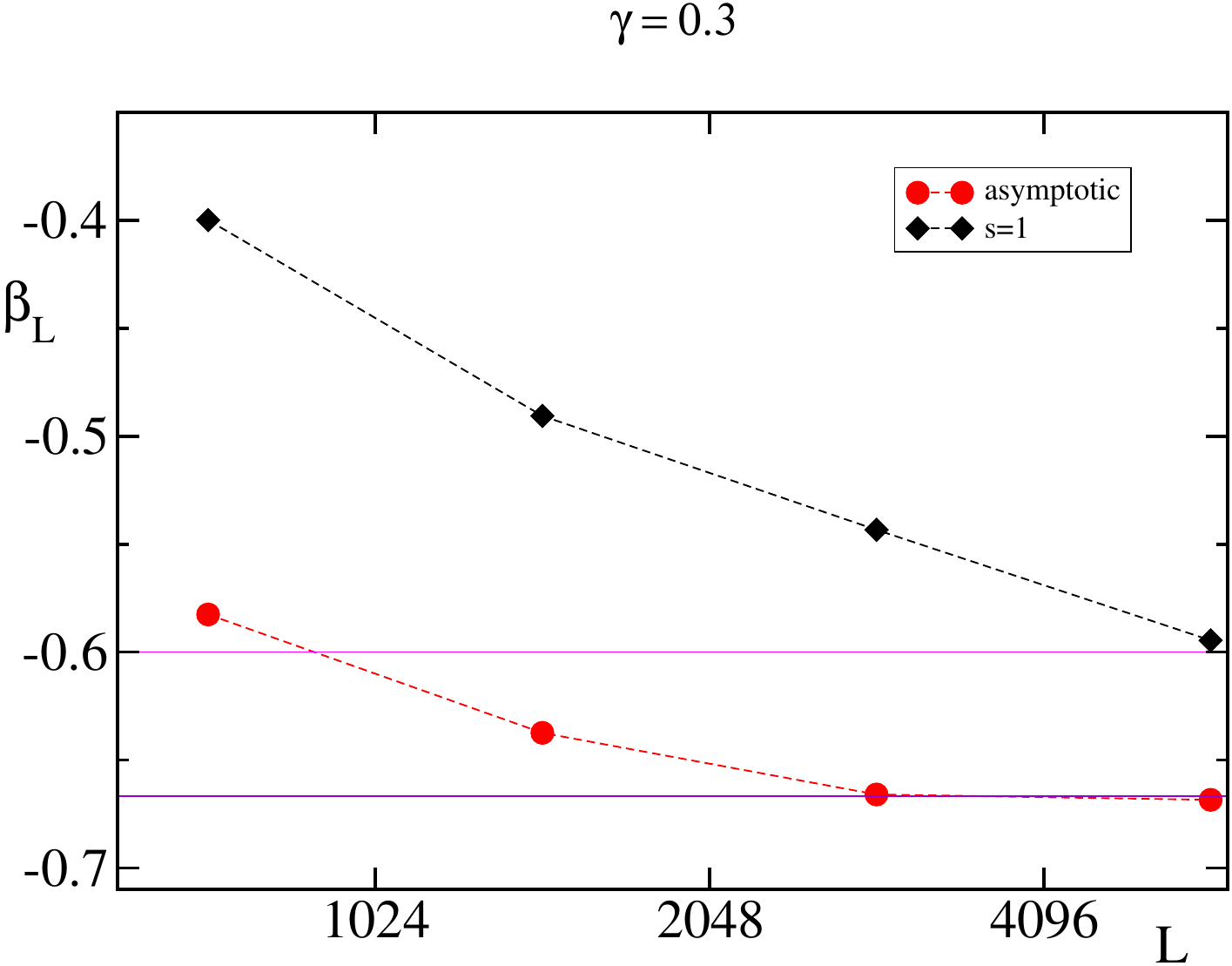}
\caption{Logarithmic derivative of the Energy flux in the FPUT-$\beta$ model. }
\label{fig:logderFPU}
\end{figure}

As a second qualitative consideration, we notice that upon increasing the system size, the dependence of the flux 
on the thermostat length decreases, suggesting that for $L\to\infty$, the flux is independent of the way the system is thermalized,
as expected for a bulk property.
A more quantitative analysis can be performed by fitting the dependence of the flux on $s$ as follows,
\begin{equation}
J(s,L) = J_{as}(L) - \delta(L) \exp[-a(L)s]
\label{dispersion}
\end{equation}
where $J_{as}$ represents the asymptotic flux value for an infinitely long thermostat,  while $a$ 
expresses the convergence rate.
Given $J(s,L)$, its logarithmic derivative $\beta$ (with respect to the chain length $L$)
yields the effective scaling behavior that is expected to converge to $ 1 - \delta = -3/5$ for the
FPUT-$\beta$ model.
In Fig.~\ref{fig:logderFPU}, we report the logarithmic derivative for the two extreme cases:
$s=1$ (black diamonds) and $s=\infty$ (red dots).
There we see large differences in the range of lengths explored therein; so large that
the asymptotic data are actually closer to the values expected for the broader universality
class (see the lower horizontal curve).
This is not a surprise, since the same strong deviations (if not even larger) had been 
observed in~\cite{takatsu2024large} (see also Fig.~\ref{fig:longbeta})
where much longer lengths have been studied to see a convergence to -0.6.
Interestingly, it looks that single-particle thermostats provide better finite-size estimate.

\begin{figure}[ht!]
\centering \includegraphics [width=0.6\textwidth,clip]{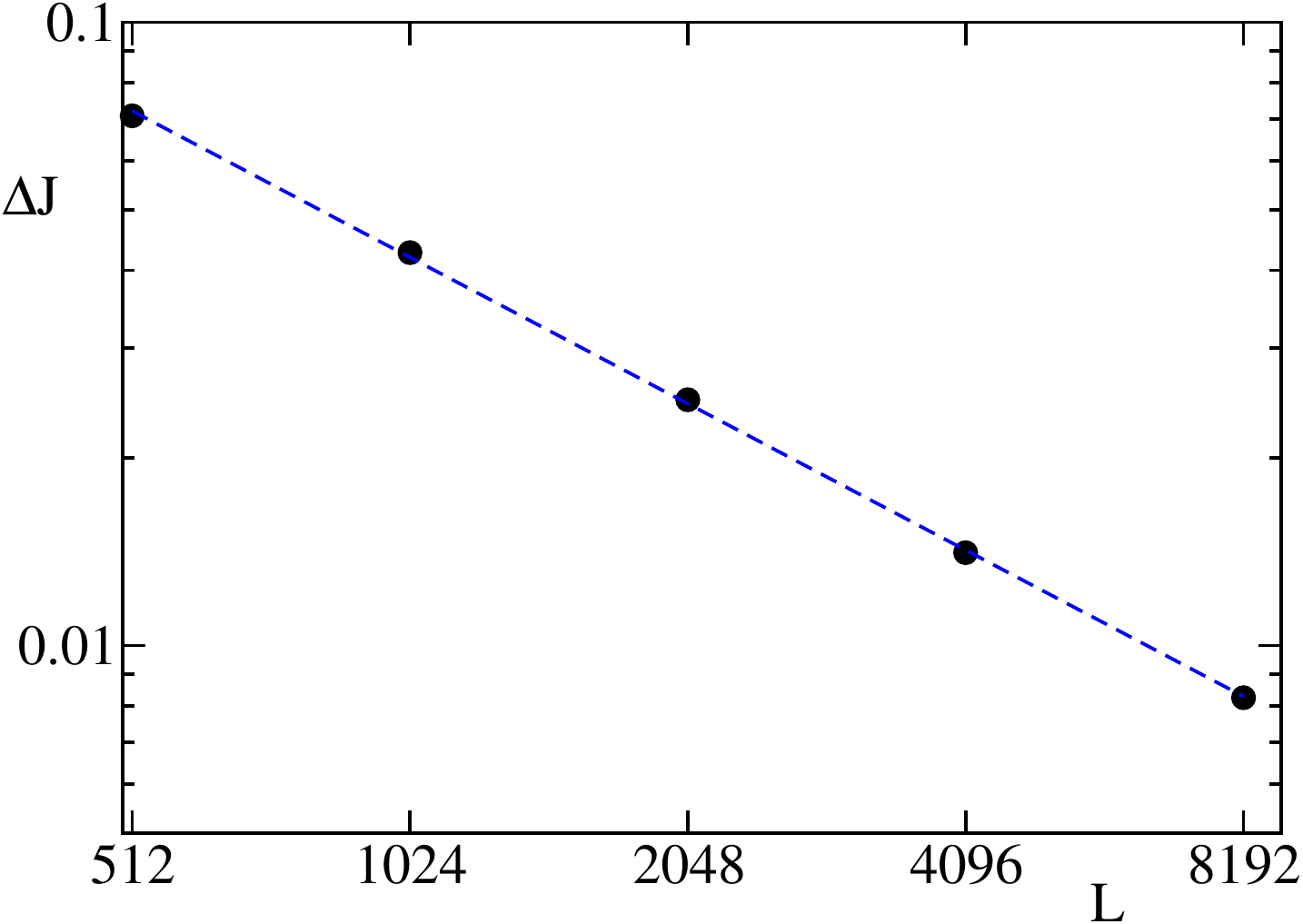}
\caption{Flux dispersion among thermostats as a function of the chain 
length $L$, see the main text and 
Eq.(\ref{dispersion}) for the definition.}
\label{fig:deltaj}
\end{figure}

Finally, we have explored the way the dispersion of fluxes among the various thermostatting lengths decreases with $L$. The results are
shown in Fig.~\ref{fig:deltaj}, where we plot $\Delta J = J_{as}(L)-J(1,L)$. We see that $\Delta J$ decreases as power law,
approximately as $L^{-0.78}$. Since the exponent is larger (although not much) than the one controlling the scaling of the
flux itself, we can conclude that the relative dispersion decays (approximately as $L^{-0.15}$) confirming the visual impression
that, eventually, the selection of the thermostat length $s$ does not matter.
However, given the finiteness of the chain length accessible in standard simulations, our results suggest that $s=1$ is the optimal choice.

\subsection{Coupling strength}

In the previous subsection, we focused on the role of the length of the thermostatted region. Here, we investigate
the implications of the  coupling strength.
We have already stated that a very weak (or strong) coupling \textit{de facto} decreases the effective temperature gradient
by inducing jumps at the borders as a form of contact resistance. In practice, this means a reduced flux.
However, once again, one expects the flux,  in the thermodynamic limit, to be essentially determined by the bulk 
dynamics. Hence, the question is the way the asymptotic behavior is progressively reached.

\begin{figure}[ht!]
\centering \includegraphics [width=0.6\textwidth,clip]{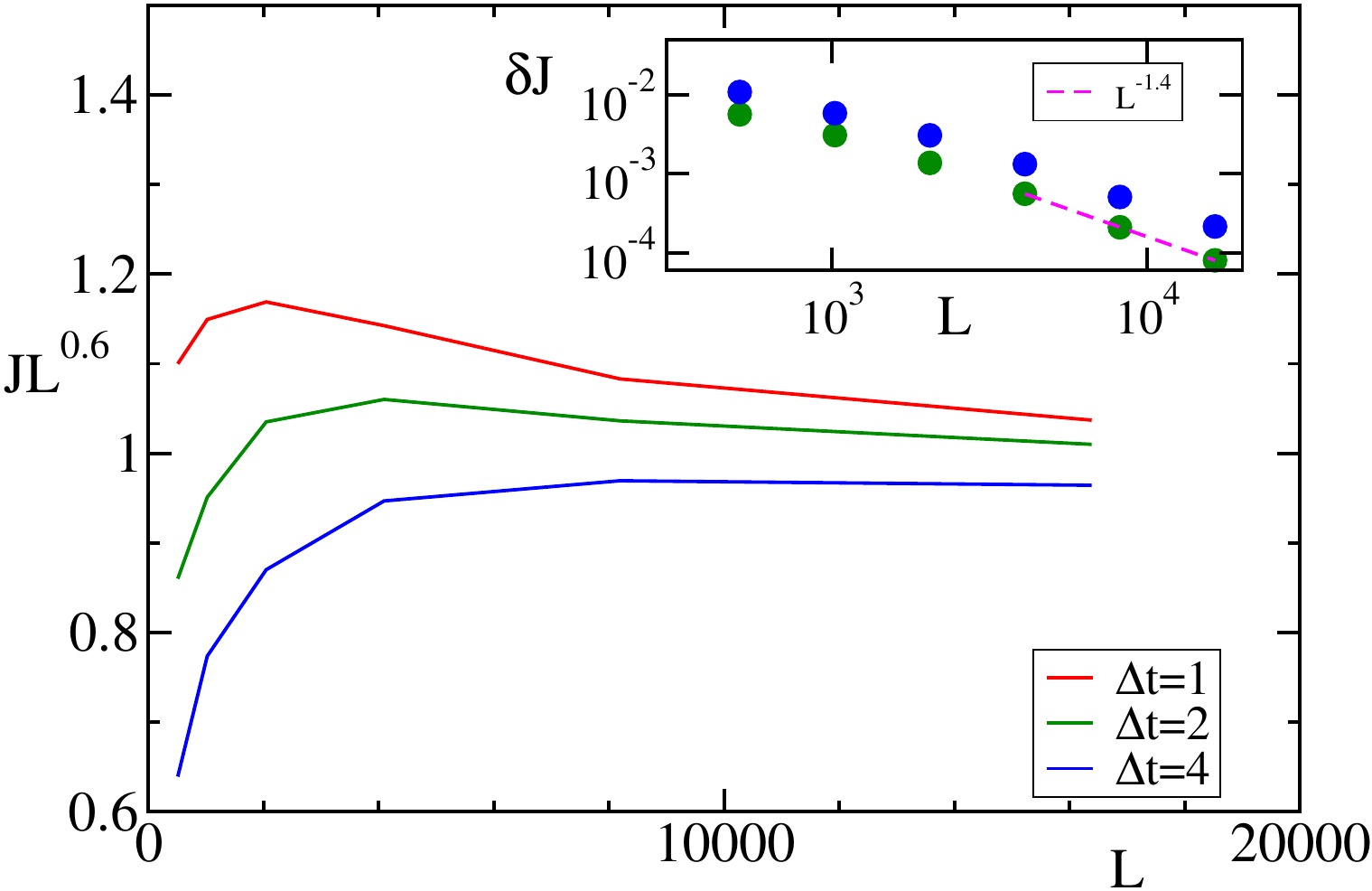}
\caption{Energy flux versus chain length in an FPUT-$\beta$ chain attached to two thermostats operating at
temperature 1.2 and 0.8, respectively. The three curves correspond to three different interaction times.
(inset) difference between the flux for $\Delta t=1$ and, respectively $\Delta t=2$ (lower curve) and 
4 (upper curve).}
\label{fig:fluxes_FPUT}
\end{figure}

In this case, the baths are simulated as random reassignments of the velocity of the 
thermostatted particle and we vary the time separation  $\Delta t $, increasing it from 1, to 2 and 4.
The results reported in Fig.~\ref{fig:fluxes_FPUT} show that the flux progressively decreases,
consistently with the expectation that a decreased strength reduces the flux
(incidentally, we plot $JL^{0.6}$ both to emphasize the expected scaling and to judge
directly the behavior of relative differences).
In the inset, we plot the difference between the maximal flux (that for $\Delta t =1$)
and the other two options, to quantify the convergence to a unique value.
The resulting scaling exponent  is about  $-1.4$, that is much faster than in the previous case.
This leads us to conclude that the choice of the coupling strength is much less critical than 
that of the thermostat length.

\section{Equilibrium simulations}
\label{sec:equil}

\subsection{Correlation functions}

Correlation functions are of basic importance to assess the nature
of transport and relaxation close to equilibrium \cite{Zhao06}.
Here, we investigate the behavior of some of them
as obtained from microcanonical equilibrium simulations.
In fact, nonlinear fluctuating hydrodynamics (see Refs. \cite{Spohn2014,VanBeijeren2012,spohn2016lnp} for details) allows us 
to identify the proper stochastic variables, characterizing the dynamics of a system with three conservation laws, as is the case for for the FPUT-chains.
They are three stochastic fields (or modes): two \textit{sound modes}, $\phi_\pm$, 
traveling at the speed of sound $c$ in opposite directions and the  \textit{heat mode} $\phi_0$, which is stationary but decaying in time.
Accordingly, the quantities of interest are the equilibrium spatiotemporal correlation 
functions $C_{s s'}(x,t)=\langle \phi_s(x,t) \phi_{s'}(0,0)\rangle$, where $s,s'=-,0,+$. 

In the generic case, which corresponds either to asymmetric inter-particle potentials or to an externally applied stress, 
the theory predicts the following scaling form for the auto-correlation functions of the modes  \cite{Spohn2014,spohn2016lnp}
\begin{eqnarray} 
  C_{\mp\mp}(x,t)&=   \frac{1}{(\lambda_s t)^{2/3}}~ f_{\mathrm{KPZ}} 
  \left[~ \frac{x \pm ct}{(\lambda_s t)^{2/3}}~\right]~, \label{eqscalS}\\
  C_{00}(x,t) &=  \frac{1}{(\lambda_e t)^{3/5}} ~f^{5/3}_{\mathrm{LW}}\left[~ \frac{x}{(\lambda_e t)^{3/5}}~\right]~.\label{eqscalE} 
\end{eqnarray}
Remarkably, the scaling function $f_{\rm KPZ}$ is universal and known exactly (see \cite{Spohn2014} and references therein). 
Also, $f_{\mathrm{LW}}^\nu(x)$ denotes the L\'{e}vy function
of index $\nu$, and $\lambda_s$ and $\lambda_e$ are  model-dependent parameters.
The above predictions have been successfully tested for several examples~\cite{Mendl2013,Das2014,Mendl2014}, including the 
FPUT-$\alpha \beta$ and related models \cite{Das2014a,DiCintio2015,Barreto2019} (although 
deviations have been reported for hard-point gas models \cite{Hurtado2016}).

Although the KPZ class is expected to be generic, there may be specific cases that belong to a different (non-KPZ) universality 
class by virtue of additional symmetries. This is the case of anharmonic chains models with symmetric interaction potentials like 
the FPUT-$\beta$ model \cite{Lepri03,Lee-Dadswell2015}.
In this case, the mode-coupling approximation of 
nonlinear fluctuating hydrodynamics predicts instead \cite{Spohn2014}:
\begin{eqnarray} 
 C_{\mp\mp}(x,t)&= \frac{1}{{(\lambda^0_s t)}^{1/2}} f_{\mathrm{G}} \left[~ 
 \frac{x \pm ct}{(\lambda_s^0 t)^{1/2}}~\right]~,
 \label{eqscalSP0}\\
  C_{00}(x,t) &=  \frac{1}{(\lambda_e^0 t)^{2/3}} ~f^{3/2}_{\mathrm{LW}}\left[~ \frac{x}{(\lambda_e^0 t)^{2/3}}~\right]~,\label{eqscalEP0}
\end{eqnarray} 
where $f_\mathrm{G}(x)$ is the unit Gaussian with zero mean. Also in this case $\lambda^0_s$ and 
$\lambda_e^0$ are  model-dependent parameters. This class would correspond to a diverging finite-size conductivity with $\delta=1/2$.

The predictions (\ref{eqscalSP0}) and (\ref{eqscalEP0}) are inconsistent with numerical simulations.
To illustrate this statement   
we consider the local energy density defined by Eq.~(\ref{eq:localv})
whose autocorrelation function 
is defined as
\[
C_{ee}(j,\tau) = \langle e_i(t) e_{i+j}(t+\tau) \rangle  - \langle e_i \rangle^2
\]
where the angular brackets denote an average over the space index $i$ and over the time $t$. Due to heat mode diffusion, 
the peak $C_{ee}(0,\tau)$ is expected to decay as $1/\tau^\gamma$. 
A well known formula links $\gamma$ with the
exponent $\delta$, which describes the divergence rate of the heat conductivity \cite{Cipriani05},
\begin{equation}
\gamma = \frac{1}{2 -  \delta}
\label{eq:alphagamma}
\end{equation}
As seen from Eq. (\ref{eqscalSP0}), fluctuating hydrodynamics predicts $\delta = 1/2$, which implies
$\gamma = 2/3$.

We have simulated the FPUT-$\beta$ model for an energy density 0.66, which corresponds to a temperature 0.75.
Averages are performed over $5\cdot 10^6$ time units.

\begin{figure}[ht!]
\centering \includegraphics [width=0.6\textwidth,clip]{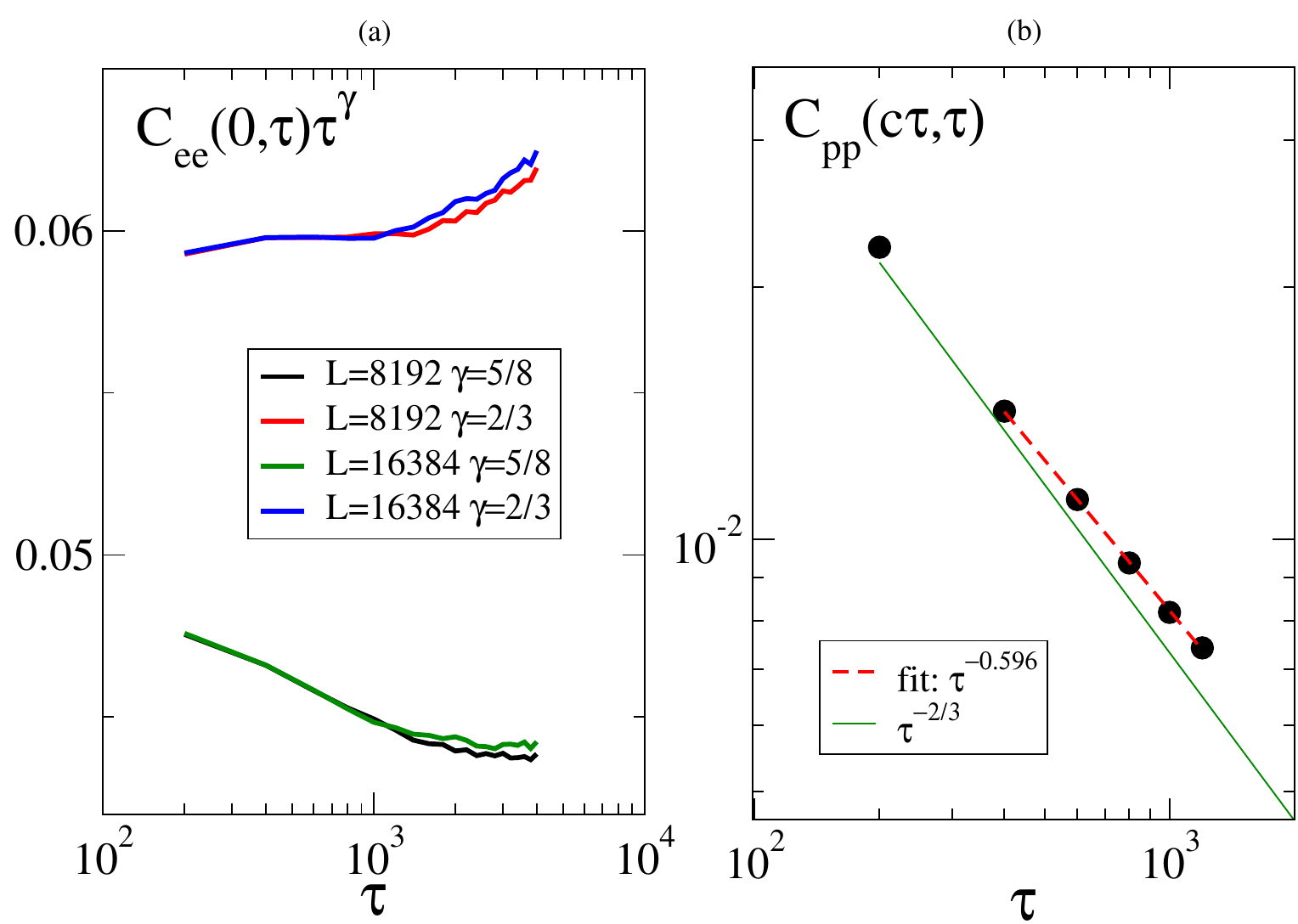}
\caption{(a) Peak of the energy autocorrelation function in FPUT-$\beta$ model. The value is averaged over 5 central points.
(b) Peak of the momentum autocorrelation function in FPUT-$\beta$ model. The peak is located in a position 
which moves with sound velocity ($\approx 1.33$ in this case). Data are obtained for a chain of length 4096; computation
are and must be stopped before  the two sound peaks reach each other. Values are averaged over 5 consecutive  points.
}
\label{fig:corr_ene}
\end{figure}

The data reported in Fig.~\ref{fig:corr_ene}a show that over long times, $\delta$ is smaller than $\frac{2}{3}$,
since the two upper curves eventually grow. The value $5/8$ is more plausible, as visible from the
behavior of the two bottom curves. This value of $\gamma$ is indeed consistent with $\delta = 2/5$, as from
Eq.~(\ref{eq:alphagamma}).

We conclude this section, by presenting the data for the decay of the sound peaks, i.e. the peaks of the autocorrelation
of the linear momentum
\[
C_{pp}(j,\tau) = \langle p_i(t) p_{i+j}(t+\tau) \rangle  - \langle p_i \rangle^2
\]
Fluctuating hydrodynamics predicts a decay of the peaks 
located at $j\approx \pm c\tau$ as $1/\tau^\frac12$, see 
Eq. (\ref{eqscalSP0}). 
In Fig.~\ref{fig:corr_ene}b we see that this is definitely out of question:
2/3 is much too large;  while 3/5 is more likely to be the right value.

Altogether, these simulations confirm that the case of symmetric potentials belongs to a universality
class different from that of KPZ equation, but different also from the prediction of fluctuating hydrodynamics and,
in agreement, with nonequilibrium stationary simulations.

\subsection{FPUT as an interface problem}

The repeatedly mentioned existence of two universality classes is deeply connected with the scenario
observed in surface roughening, described by the celebrated Kardar-Parisi-Zhang (KPZ) equation, 
which, in the limit of a vanishing interface velocity, reduces to the Edwards-Wilkinson model.
As mentioned in the previous subsection,
nonlinear fluctuating hydrodynamics  \cite{Spohn2014,VanBeijeren2012,spohn2016lnp} has allowed unveiling this relationship by showing that,
at a mesoscopic level, the dynamics of a nonlinear chain, obeying the three standard conservation laws, is 
described by three stochastic fields or modes, namely the two \textit{sound modes}, $\phi_\pm$, 
and the  \textit{heat mode} $\phi_0$.
Loosely speaking, the dynamics of each sound mode $\phi_\pm$ follows a fluctuating Burgers equation in a moving frame,
\begin{equation}
\frac{\partial \phi_\pm}{\partial t}=\pm c  \frac{\partial \phi_\pm}{\partial x} + 
\lambda \frac{\partial \phi_\pm^2}{\partial x} + 
D\frac{\partial^2 \phi_\pm}{\partial x^2} + \frac{\partial \eta_\pm}{\partial x},
\label{burgers}
\end{equation}
where $\eta$ is a white noise.

Following linear hydrodynamics, a sound mode is (in the zero pressure case) a linear combination of
stretch and  momentum
 \begin{equation}
\phi_\pm(n) = \pm c r_n + p_n
\label{rhoFPU}
\end{equation}
The Burgers equation can be straightforwardly transformed into a KPZ equation, by introducing the 
interface height $h_\pm = \int dx \phi_\pm$,
\begin{equation}
\frac{\partial h_\pm}{\partial t}= \pm c \frac{\partial h_\pm}{\partial x} + 
\frac{\lambda}{2} \left(\frac{\partial h_\pm}{\partial x}\right)^2 +D\frac{\partial^2 h_\pm}{\partial x^2} +  \eta \; .
\label{kpzh}
\end{equation}

For the sake of numerical computation, me define the microscopic version of the interface height as
\begin{equation}
h_\pm(n) = \pm c q_n + \sum_{j=1}^n p_j \; .
\label{var}
\end{equation}
The underlying roughening is studied by monitoring the evolution of the interface width
\begin{equation}
W = \sqrt{\langle h_\pm^2 \rangle - \langle h_\pm\rangle^2}
\label{hwidth}
\nonumber
\end{equation}
starting from the initial condition $W(0)=0$. 
Here, we cannot select a strictly flat initial profile, since a perfect initial spatial homogeneity 
would persist at all times (the noise is self-generated).
In practice, we first select the microcanonical temperature $T$ and then identify the corresponding
potential energy density $U$. Next, we proceed by randomly choosing
the $q_i$'s according to a Gaussian distribution\footnote{This is not optimal, but local equilibrium is
nevertheless rapidly attained.} in such a way that the potential energy density is indeed $U$.
As for momenta initialization, the procedure is a bit more complicated: the definition of the interface (\ref{var}) involves
the integral $\tilde p_n = \sum_i^n p_j$, which we want to be uncorrelated among themselves
(to ensure the initial profile to be as flat as possible). 
In practice, the $\tilde p_n$'s are randomly chosen according to a Gaussian a distribution, in such
a way that the kinetic energy density (defined in terms of the $p_n$ values) is equal to $T/2$.

\begin{figure}[ht!]
\centering \includegraphics [width=0.75\textwidth,clip]{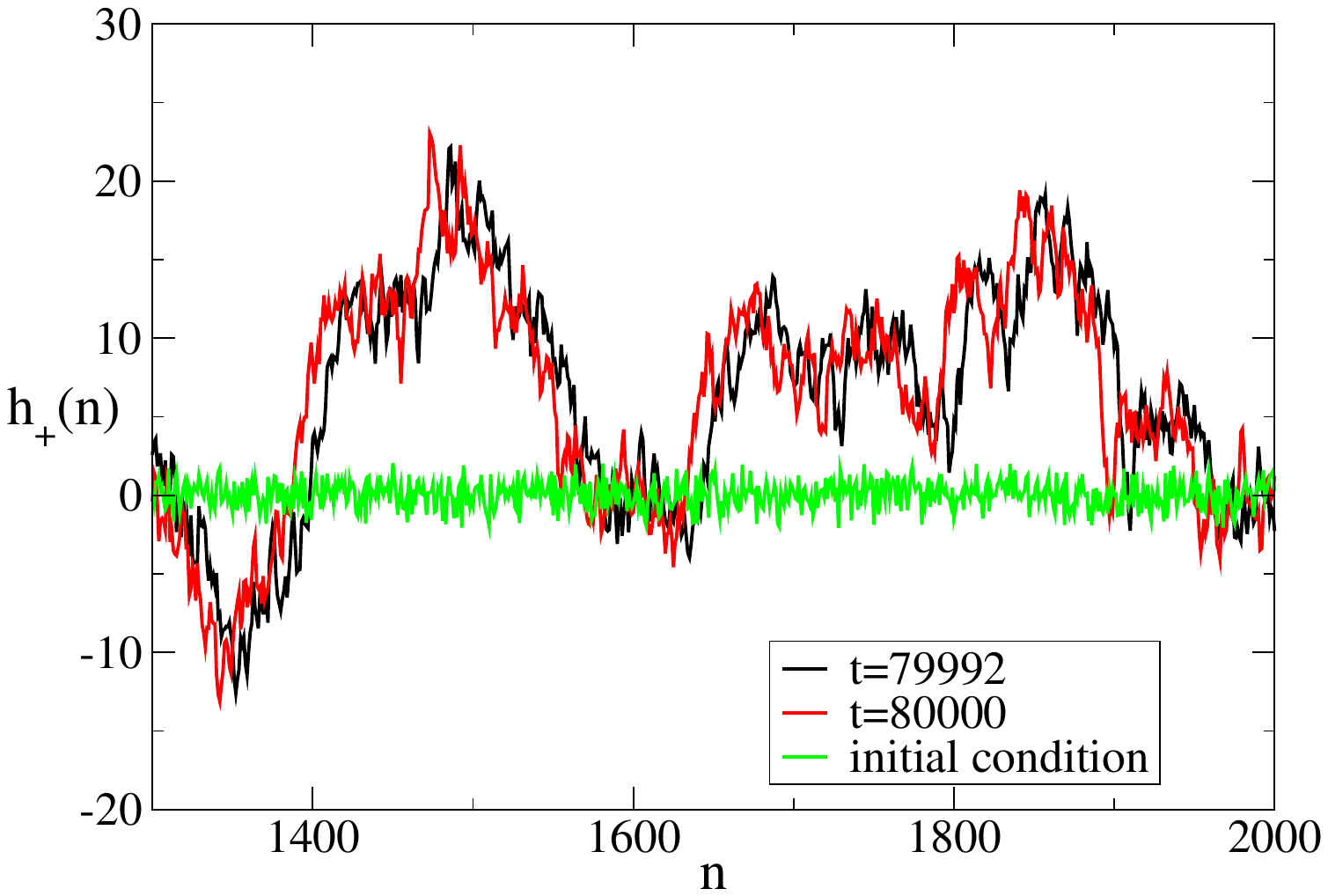}
\caption{Samples of interface profiles as defined in 
Eq.(\ref{var}) for the FPUT-$\beta$ model at temperature $T=0.75$
ad two subsequent times.
The initial density of potential energy is $U = 0.2838$; sound velocity is 
$c= 1.3955$.}
\label{fig:interface_b}
\end{figure}

The green line in Fig.~\ref{fig:interface_b} is a section of a typical initial condition: in spite of
the unavoidable local fluctuations, the interface is significantly flat.
In the same figure we report two additional instances of the same sound-mode profile,  sampled after a long time 
separated by a time interval of 8 units: the presence of a drift is clearly visible.

\begin{figure}[ht!]
\centering \includegraphics [width=0.75\textwidth,clip]{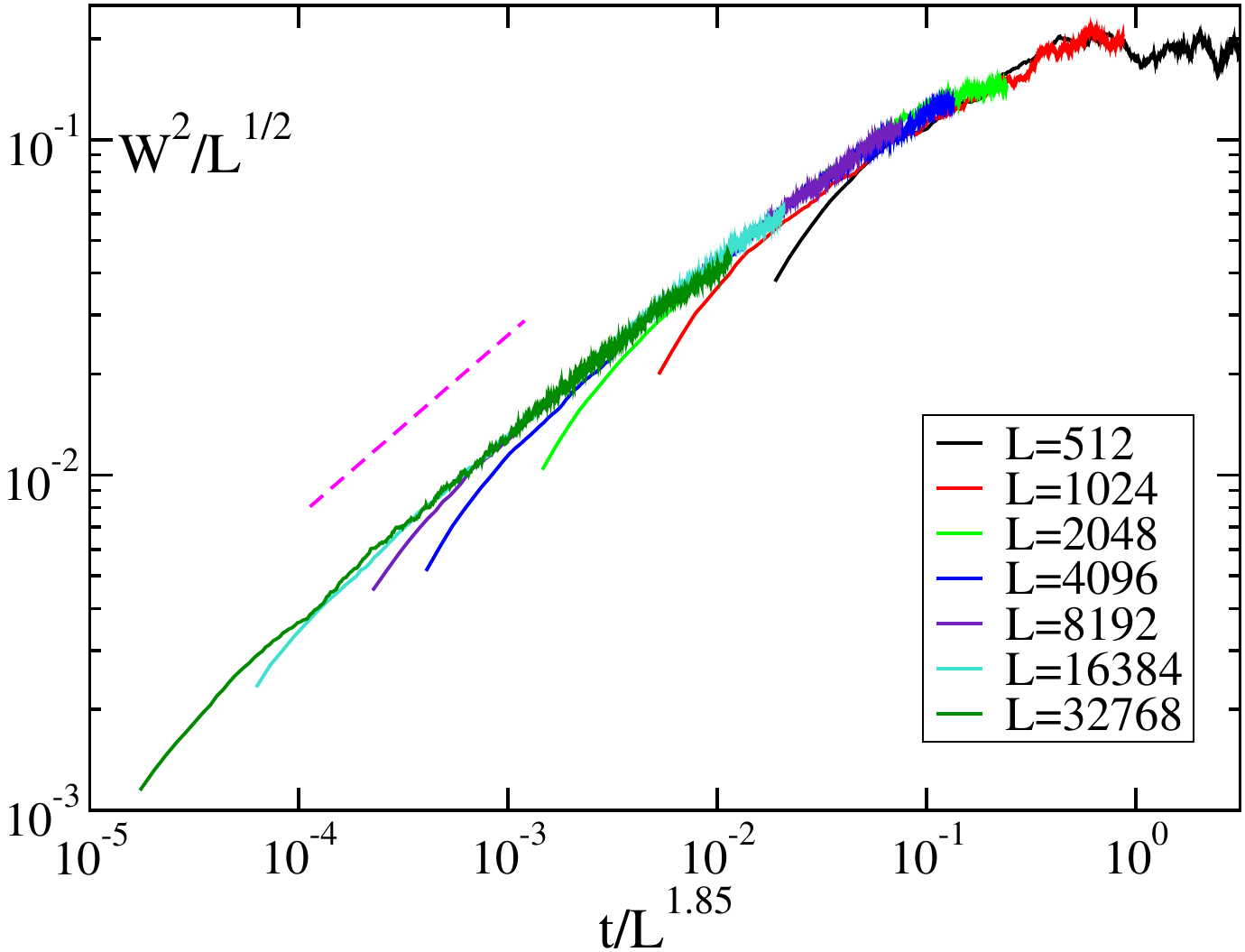}
\caption{Interface width for FPUT-$\beta$ chains at temperature $T=0.75$.
The initial potential energy is $U = 0.2838$; sound velocity is $| c | =  1.3955$.
Data are obtained by averaging over 100 realizations.
The dashed line corresponds to a slope $0.27$.}
\label{fig:interface_ab}
\end{figure}

A quantitative analysis of the interface dynamics for the FPUT-$\beta$ model is presented in
Fig.~\ref{fig:interface_ab} where both axes have been suitably scaled to ensure an optimal data
collapse\footnote{Because of the periodic b.c. adopted in the simulations, 
the presence of the drift does not affect the dynamics of the width.}

In the presence of a dynamical scaling, one expects $W = L^\alpha G(t/x^z)$ and $G(u)\approx u^\beta$ for $u\ll 1$
for some values of the various exponents \footnote{Here we keep the standard notations adopted in the literature for
the scaling exponents $\alpha$, $\beta$ and $z$ of the interface width $W$. Unfortunately this might generate some
misunderstanding, because the greek letters $\alpha$ and $\beta$ are also adopted in the literature for identifying the 
two kinds FUPT-chains. Manifestly, this is just due to chance.}. 
Whenever Galilean invariance is satisfied, as in the present case,
we also know that $\beta = \alpha/z$~\cite{barabasi1995fractal}. 
In the KPZ case, $\alpha=1/2$, $z=3/2$, and $\beta=1/3$, while for the Edwards-Wilkinson (EW)
model, $\alpha=1/2$, $z=2$, and $\beta=1/4$~\cite{barabasi1995fractal}.

From the data reported in Fig.~\ref{fig:interface_ab}, there is a perfect agreement for the $\alpha$ value,
which, by the way, reflects the simple fact that the asymptotic profile is a standard Brownian bridge.
The $z$ exponent turns out to lie in between the value expected for the EW (2) and the KPZ ($3/2$) model.
Furthermore, a direct fit of the power-law growth of the width yields an exponent $\approx 0.27$ (see the dashed line),
in agreement with the expectation from Galilean invariance $1/(2 \times 1.85)$, and again in between  
EW (1/4) and KPZ (1/3) expectations.
More quantitatively, coherence with the equilibrium simulations discussed in the previous sub-section
would require a slightly lower $z=5/3$ and larger $\beta = 3/10$.
Altogether, a result emerges from these interface-like simulations: 
the FPUT-$\beta$ model differs from both the KPZ and EW dynamics, 
suggesting that it belongs to a different and yet unknown universality class.

\section{Adding conservative noise}
\label{consnoise}

In view of the difficulty of dealing with genuine nonlinear interactions, a
class of hybrid dynamics, where a deterministic evolution is 
accompanied by stochastic interactions has been proposed \cite{BBO06,Basile08,basile2016thermal}. A basic requirement is that the 
stochastic process preserves the conservation
laws, Eqs. (\ref{eq:localv}).   
This is referred to as \textit{conservative noise} dynamics: in its simplest versions
it amounts to considering a simple harmonic chain with 
random exchange of momenta between a couple on neighbouring 
particles. This model allows sometimes for exact solutions \cite{Lepri2009,Lepri10,Delfini10,Kundu2019} and can be simulated 
by a numerically exact algorithm
\cite{lepri2023thermalization}.

One may also wonder about the effect of conservative noise on nonlinear
oscillator chains \cite{Basile08,Iacobucci2010,Bernardin2014,lepri2020too}.
In the case of FPUT dynamics, some  
previous numerical simulations suggested that the scaling behavior bends towards the $-1/2$ exponent \cite{Basile08}.  
(as in the harmonic model with conservative noise \cite{BBO06}).

We have revisited the numerical analysis, since in the light of the fluctuating-hydrodynamics theory, it is
unclear the reason why the scaling behavior should be accelerated by the addition of an extra conservative
noise, which, a priori, should more reasonably lower the conductivity (increase the thermal resistance).

\begin{figure}[ht!]
\centering \includegraphics [width=0.7\textwidth,clip]{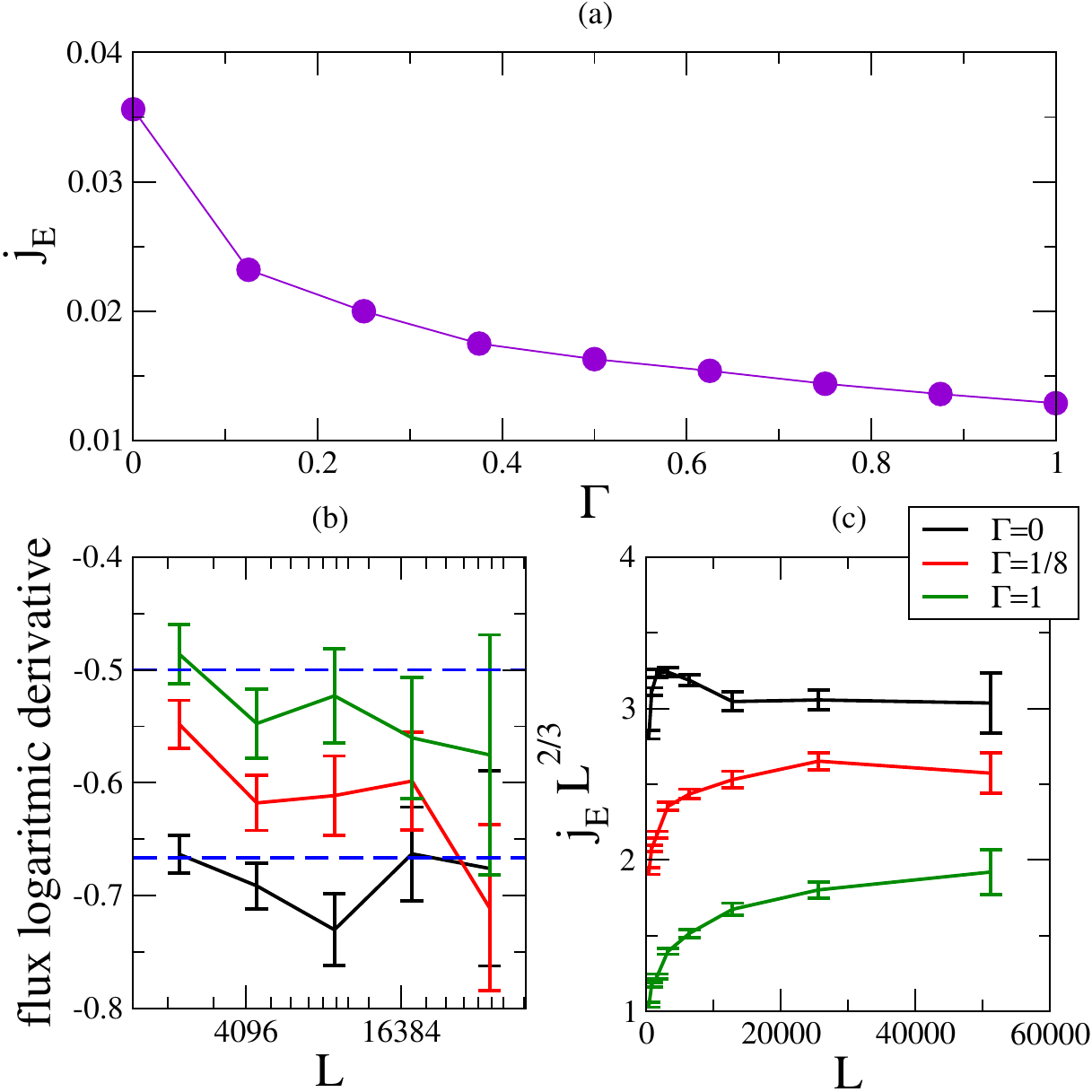}
\caption{Simulations of FPUT-$\alpha \beta$ with 
conservative noise whereby the momenta of two neighbouring particles 
are exchanged with a fixed rate $\Gamma$; (a) Energy flux for a given length $L=800$ and different $\Gamma$ rates of the conservative noise. 
(b) Logarithmic derivative of the flux; (c) flux scaled by a factor $L^{2/3}$ }
\label{fig:flux_gamma}
\end{figure}

We have simulated the FPUT-$\alpha \beta$ model with equal coefficients 
of two nonlinearities and with thermal baths at temperatures 4.5, 5.5. 
Conservative noise is implemented by exchanging the momenta
of pair of neighbouring particles, $(p_n,p_{n+1})\to(p_{n+1},p_n)$ 
with a given rate $\Gamma$ \cite{Basile08}. 
The data reported in Fig.~\ref{fig:flux_gamma}a 
for  different values of $\Gamma$ ($\Gamma=0$ corresponding  to the deterministic system) 
show that, indeed, increasing the noise amplitude yields
a decrease of the flux, as expected.

As for the scaling dependence of the flux, since it requires very long simulations, we focused on three
specific values, $\Gamma=0$ (deterministic), $\Gamma=1/8$ and $\Gamma=1$.
The results are reported in Fig.~\ref{fig:flux_gamma} by adopting two different approaches.
In Fig.~\ref{fig:flux_gamma}b , we report the logarithmic derivative, computed by comparing the flux for two different
chain lengths. The derivative is very sensitive to noise (the error has been estimated by summing the
contribution of the uncertainty on both fluxes in each difference). Anyhow, one can qualitatively
appreciate that the scaling behavior tends to depart from the case of purely harmonic chain with conservative noise $\delta=1/2$ \cite{BBO06}
and rather
to (slowly) converge towards the hydrodynamic prediction
$\delta=1/3$. This conclusion is perhaps more transparent
in Fig. \ref{fig:flux_gamma}c, where the flux is scaled by the theoretical prediction.
As a final statement, we consider conceptually unlikely that either the red or the green curve
eventually cross the black deterministic curve, and conclude that addition of conservative
noise does not change the universality class.

\section{Transport close to the integrable limit}
\label{sec:transint}

Almost-integrability effects are responsible of many 
features of relaxation in the classic FPUT problem.
It is thus relevant to investigate how they affect transport. 
The above results are mostly obtained in strongly nonlinear regimes, far from any integrable limit. For the FPUT models, this means working with high enough energies/temperatures.

Integrable systems constitute \textit{per se} a relevant case and experienced a renovated interest in recent years.  In the framework of the present work, the main example is certainly the celebrated 
Toda chain, namely the model in Eq.~(\ref{hami}) with  
$V=V_T(x) = e^{-x} + x -1$. This integrable model exhibits ballistic heat transport, 
as first suggested by Toda himself \cite{Toda79}. 
Heuristically, it means that heat-carriers are 
the quasi-particles (the famous Toda solitons) that experience a stochastic sequence of spatial shifts as they move through the lattice, interacting with other excitations without momentum exchange.  One way to visualize such quasi
particles is to look at eigenvectors of the Lax matrix \cite{lepri2025lax}.

Ballistic transport manifests  
by a non-zero \textit{Drude weight} namely, a zero-frequency component of  energy-current power spectra \cite{Zotos02,Shastry2010} or, 
equivalently, by the fact that flux autocorrelation does not 
decay to zero at large times. Also,   
while transport is predominantly ballistic, the non-dissipative interactions introduce a minor, finite diffusive component \cite{Shastry2010,Kundu2016,DiCintio2018}.

For a generic perturbation of the Toda chain, still conserving the three basic quantities 
(stretch, momentum and energy) one expects a change from ballistic 
to anomalous conductivity, for large enough system sizes. 
A relevant issue regards the typical length scales over which the anomalous transport sets in. The length-independent flux exhibited by integrable systems is the result of the free displacement of quasi-particles (the integrals of motion, such as solitons) from the hot towards the cold reservoir.
In the vicinity of the integrable limit, as a result of mutual interactions, the quasi-particles acquire a finite and large mean free path $\ell$. A purely ballistic behavior is observed for $L<\ell$. On the other hand, $L>\ell$ is not a sufficient condition to observe a crossover toward the anomalous behavior predicted by the above-mentioned theoretical arguments.
In fact, it is necessary for $L$ to be so long that the {\it normal} flux induced by inter-particle scattering becomes negligible.
Altogether, upon increasing $L$ at fixed $\ell$, one should see a first ballistic regime followed by a kinetic (diffusive) one, until 
eventually, the asymptotic hydrodynamic (anomalous) regime is attained.
The three different regimes are observable only provided the relevant length scales are widely separated.

Based on these heuristic considerations, one may look for a decomposition of heat flux $J(L,\varepsilon)$ as \cite{lepri2020too}
\begin{equation}
J(L,\varepsilon) = J_A(L,\varepsilon) + J_N(L,\varepsilon),
\label{Jtot}
\end{equation}
where $\varepsilon$ measures the perturbation strength
i.e. the distance from the integrable limit, $J_A$ is the anomalous hydrodynamic part, and
$J_N$ is the kinetic contribution, accounting for the energy transported by the weakly interacting quasi-particles. As explained above, for $L\to\infty$,
we expect $J_A \approx L^{\delta-1}$ with $\delta=1/3$ in systems belonging to the KPZ class.

Following a kinetic argument~\cite{pitaevskii2012physical}, we argue that $J_N$ must be 
only a function of  $\xi=L/\ell$, which is the ratio expressed in units 
of the mean free path $\ell$, the only relevant scale. Moreover, $J_N$ should display a crossover from ballistic to diffusive regimes depending on $\xi$, namely it should
approach a 
constant for small $\xi$ and  be proportional to $1/\xi$ for large $\xi$. A simple interpolating 
formula would thus be 
\begin{equation}
J_N(\xi) = \frac{j_0}{r+ \xi}\; ,
\label{JN}
\end{equation}
where $r$ is a constant accounting for the boundary resistance \cite{Aoki01} 
and $j_0$ is an additional constant.

Approaching the integrable limit  
the mean free path must diverge, and it is natural to assume 
that $\ell \approx \varepsilon^{-\theta}$, 
where $\theta >0$ is a system-dependent exponent.
As long as $J_A(L,\varepsilon)$ does not display any singularity for $\varepsilon \to 0$ (we return to this point below), we can neglect its dependence on
$\varepsilon$. 
Altogether, Eq.~(\ref{Jtot}) can be approximated
for large $L$ as
\begin{equation}
J(L,\varepsilon) \;\approx \; \frac{c_A}{L^{1-\delta}} \,+\, 
 \frac{c_N}{L\varepsilon^{\theta}}  \; ,
\label{Jtot2}
\end{equation}
where $c_A$ and $c_N$ are two suitable parameters. Accordingly,
the anomalous contribution dominates only above the crossover length
$\ell_c \approx \varepsilon^{-\theta/\delta}$. For $L\le \ell_c$, heat conduction is
dominated by $J_N$. In particular, within the range $[\ell=\varepsilon^{-\theta},\ell_c]$
an \textit{apparent} normal conductivity is expected, 
which is nothing but a finite size effect.
 
\begin{figure}[ht!]
\centering \includegraphics [width=0.8\textwidth,clip]{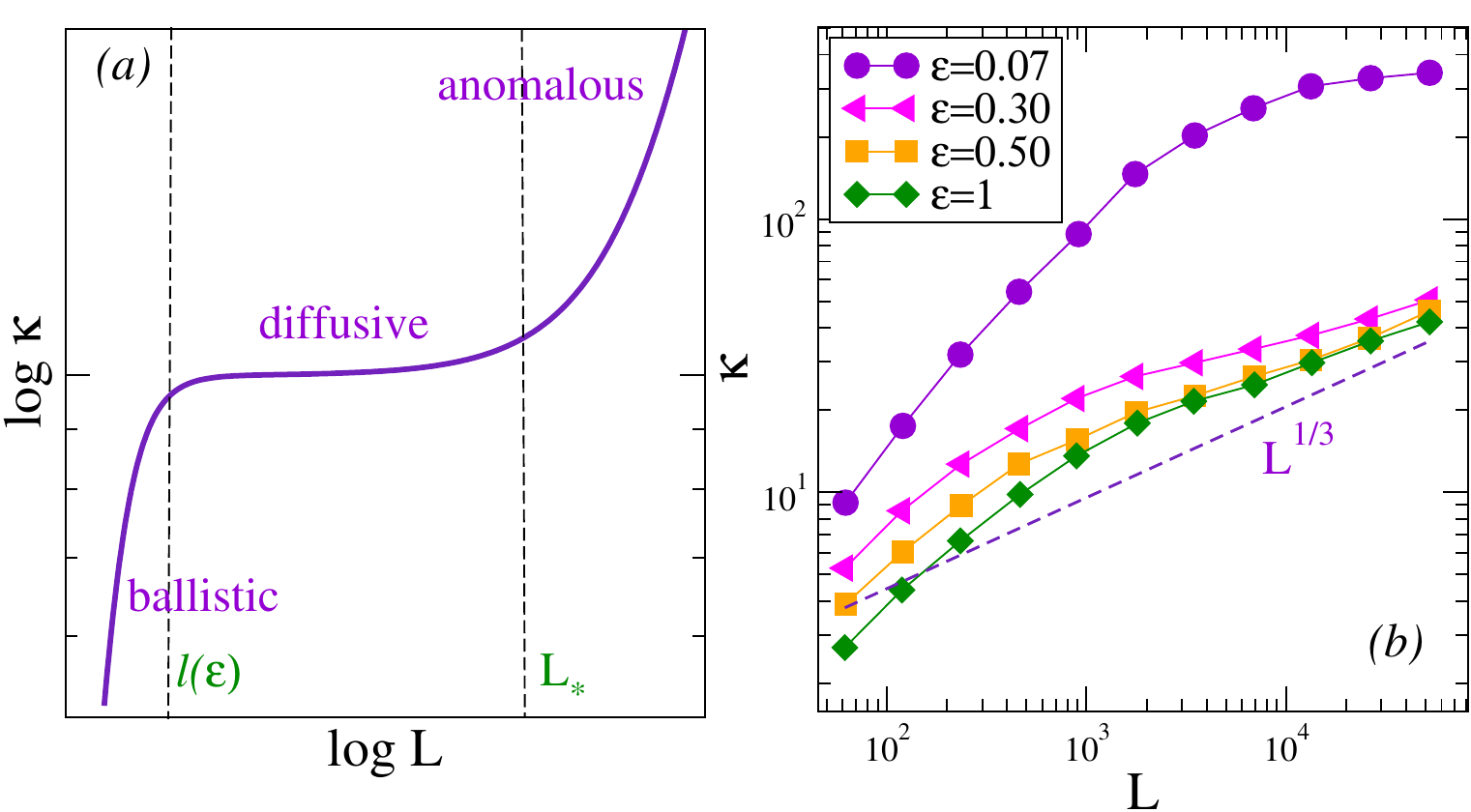}
\caption{Illustrating the effect of almost-integrability on 
heat transport: (a) scheme of three different regimes 
discussed in the text (b) numerical simulation of the 
Toda model with alternating masses $m_{1,2}$. The parameter
$\varepsilon=m_1/m_2 - 1$ denotes the strength of 
the integrability-breaking perturbation \cite{lepri2020too}.}
\label{fig:dtoda}
\end{figure}

To summarize, upon increasing the lattice size $L$ 
on should observe three different regimes (see Fig. \ref{fig:dtoda})
\begin{enumerate}
    \item \textit{Ballistic Regime ($L < \ell$):} Constant heat flux $J$.
    \item \textit{Kinetic/Diffusive Regime ($\ell < L \le \ell_c$):} Finite-size effect showing \emph{apparent normal diffusion} with $J \propto 1/L$.
    \item \textit{Anomalous Regime ($L > \ell_c$):} The asymptotic anomalous component $J_A \propto L^{\delta-1}$ dominates.
\end{enumerate}
The intermediate diffusive regime is thus a \textit{transient finite-size effect} occurring before the asymptotic KPZ scaling is reached.

The above arguments should apply also to low-temperature transport of 
the FPUT-$\alpha \beta$ model. As it is known, its dynamics 
can be better described in term of a perturbed Toda model \cite{Benettin2013}.
Simulations of the FPUT-$\alpha\beta$ 
chain are quite consistent with the above scenario \cite{Das2014}.
More generally, it can explain the numerical
observation of the apparent normal diffusion observed
for generic asymmetric potentials $V(x)\neq V(-x)$ ~\cite{Iacobucci2010,Chen2014,Zhong2012,Wang2013}. 
Indeed, if the potential is well approximated by a perturbed 
Toda one, the crossover to the anomalous regime may occur at 
prohibitively large sizes. 
Interestingly, the same type of cross-over scenario is observed  
for other one-dimensional and quasi one-dimensional models close
to integrability 
\cite{Miron2019,lepri2021kinetic,kundu2023super,luo2025multi}.
Altogether, the analysis reported in this section suggests that this mechanism
might be at the basis of the seemingly
normal conductivity observed in several other models, as indeed confirmed in Ref.~\cite{politi2025}. 

To conclude this Section, we discuss the case of FPUT-$\beta$ at low 
temperature. In this 
case the closest integrable model is the harmonic chain \cite{Benettin2013}, 
quasi-particle are phonons \cite{li2010energy} and a different behavior is expected.
Indeed, the intermediate diffusive regime is absent \cite{Lepri05}. This can be 
explained by formula (\ref{Jtot2})
by a  singular dependence of the coefficient controlling 
the anomalous contribution, $c_A \approx \varepsilon^{-1/2}$, which causes $J_A$ to dominate $J_N$ for all $L$, yielding a direct ballistic-to-anomalous crossover . 

\section{Conclusions}
In this review paper we have analyzed the FPUT model, showing that,
depending on the potential symmetry, its behavior can fall within two
distinct universality classes, thereby
confirming that it is a proper testbed for the study of heat conduction in one dimension.
On the one hand, we reiterate that, as predicted by fluctuating hydrodynamics, the
the divergence of heat conductivity in the FPUT-$\alpha \beta$ model falls within 
a broad class, which can be traced back to the KPZ dynamics of rough interfaces.
On the other hand, while confirming that the FPU-$\beta$ model does belong
to a different universality class, as correctly anticipated by fluctuating hydrodynamics,
we show that its behavior differs from the expected analogy with the Edwards-Wilkinson model,
while agrees with the prediction of kinetic theory.
Moreover, our equilibrium simulations provide direct evidence
that the interface dynamics associated to the FPUT-$\beta$ model belongs to a further
yet unknown universality class.

Furthermore, a great deal of this review is devoted to a discussion of how finite-size
effects may affect the hydrodynamic behavior of such models. Far for being an academic exercise,
these studies are of primary importance for comparing theoretical studies with experimental 
tests. In fact, nanotechnologies have quite recently made possible the study of heat transport in nanotubes,
polymers, spin chains and graphene layers where size effects are much relevant
\cite{zhang2020size}. 
Lacking a careful control of them in any concrete
experimental setup, it is hopeless to check the validity of asymptotic hydrodynamic predictions, as well
as to identify and quantify possible finite-size effects.

Finally, there is a number of possible new effects that may arise in 
nonequilibrium FPUT dynamics. We just mention the role of long-range
forces, notably coupling which decay as inverse power of the distance 
\cite{Bagchi2017,Iubini2018} and the interplay between disorder
and nonlinearity  \cite{Dhar2008,zhu2021effects}. 

\vspace{10pt}
\noindent\textbf{Acknowledgements:} 
SL aknowledges support from the Italian MUR
PRIN2022 project “Breakdown of ergodicity in classical
and quantum many-body systems” (BECQuMB) Grant
No. 20222BHC9Z. 

\vspace{10pt}
\noindent\textbf{Data availability:} The datasets generated during and/or 
analysed during the current study are available from the 
corresponding author on reasonable request.

\vspace{1cm}

\section*{References}
\bibliography{bibliofput}
\end{document}